\documentclass[journal=nalefd,manuscript=communication,layout=twocolumn]{achemso}
\usepackage[version=3]{mhchem}
\usepackage{amsmath,amssymb,amsfonts}
\usepackage{color}
\usepackage{graphicx}
\usepackage{dcolumn}
\usepackage{bm}
\usepackage{color,soul}

\usepackage[usenames]{xcolor}

\author{Mingkai Liu}
\affiliation {Nonlinear Physics Centre, Research School of Physics and Engineering, Australian National University, Canberra ACT 2601, Australia}
\altaffiliation{Equal contribution}
\email{mingkai.liu@anu.edu.au}

\author{Duk-Yong Choi}
\affiliation {Laser Physics Centre, Research School of Physics and Engineering, Australian National University, Canberra ACT 2601, Australia}
\alsoaffiliation {College of Information Science and Technology, Jinan University, Guangzhou, Guangdong 510632, China}
\altaffiliation{Equal contribution}

\keywords{metasurfaces, Huygens' condition, bound states in the continuum, all-dielectric, dispersion}

\title{Extreme Huygens' metasurfaces based on quasi-bound states in the continuum}

\begin{document}
\begin{abstract}
We introduce the concept and a generic approach to realize \emph{Extreme Huygens' Metasurfaces} by bridging the concepts of Huygens' conditions and optical bound states in the continuum. This novel paradigm allows creating Huygens' metasurfaces whose quality factors can be tuned over orders of magnitudes, generating extremely dispersive phase modulation. We validate this concept with a proof-of-concept experiment at the near-infrared wavelengths, demonstrating all-dielectric Huygens' metasurfaces with different quality factors. Our study points out a practical route for controlling the radiative decay rate while maintaining the Huygens' condition, complementing existing Huygens' metasurfaces whose bandwidths are relatively broad and complicated to tune. This novel feature can provide new insight for various applications, including optical sensing, dispersion engineering and pulse-shaping, tunable metasurfaces, meta-devices with high spectral selectivity, and nonlinear meta-optics.   
\end{abstract}



\section{Introduction}

Recent advances in Huygens' metasurfaces and metadevices provide new insight into highly efficient wavefront shaping and scattering manipulation by incorporating resonances of different parities \cite{pfeiffer2013metamaterial,liu2010incident,monticone2013full,kim2014optical,pfeiffer2014efficient, asadchy2015broadband,epstein2016cavity,
PhysRevLett.117.256103,wong2016reflectionless,elsakka2016multifunctional,estakhri2016wave} (Figure~\ref{fig:concept}a).  
The introduction of high-index, low-loss dielectrics offers a practical route to achieve Huygens' condition at optical frequencies \cite{decker2015high,arbabi2015dielectric,chong2015polarization,kruk2016invited,wang2016grayscale,khorasaninejad2016metalenses,kruk2017functional,wang2018broadband}. So far, all realized Huygens’ metasurfaces are relatively broadband due to the strong radiative loss of the resonances excited, which is desirable for static linear devices that require a broadband operation. However, such a broadband feature also brings in great challenges when creating tunable metadevices with a large dynamic range, or nonlinear optical devices that require high nonlinear efficiency, or metasurfaces that can control the chromatic dispersion of pulses. Can we find a generic approach that can control the quality factors (Q-factors) of the resonances involved while maintaining the Huygens' condition?    

Fundamentally, the Q-factor of a non-absorbing resonant system is determined by the radiative loss of the resonant mode, which is defined by the total overlapping between the modes and the polarizations of the scattering channels. As the overlapping approaches zero, the Q-factor can increase to infinity. This novel feature is directly linked to the important concept called bound states in the continuum (BICs) \cite{hsu2016bound}. BICs are states that remain localized even though they coexist in the same spectrum with a continuum of radiating waves that can carry energy away. Being first theorized in quantum mechanics~\cite{von1929On}, BICs are found to be quite general wave phenomena and have been predicted and observed in various classical systems including photonic crystals, waveguides and even subwavelength resonant particles \cite{marinica2008bound,plotnik2011experimental,lee2012observation,molina2012surface,hsu2013observation,weimann2013compact,monticone2014embedded,gomis2017anisotropy,kodigala2017lasing,bulgakov2017light,rybin2017high}.  In practice, the Q-factor of any observable BIC is large but finite due to the loss introduced from various sources including finite sample size and fabrication imperfection, and therefore these observable BICs are all quasi-BICs. A typical feature of quasi-BICs in resonant systems is that the Q-factor of the resonance can approach infinity asymptotically when the parameter of the resonator approaches a critical point~\cite{hsu2016bound}. This phenomenon occurs when the total overlapping of the resonant modes and the radiative channels, which can be defined by the dot product of the total electric/magnetic polarization and the incident polarization, approaches zero (see the bottom plot of Figure~\ref{fig:concept}a as an example), and such singularities have been shown to possess certain topological properties \cite{zhen2014topological,bulgakov2017topological,doeleman2018experimental}. Our recent study further shows that many of the previously studied asymmetric metasurfaces that support high-Q Fano resonances (so-called ``dark-modes") actually share the same physics of BICs~\cite{Kirill2018Asymmetric}. 

Here, we show that these two seemly unrelated concepts, i.e. Huygens's condition and bound states in the continuum can be bridged via a novel metasurface platform. We introduce a generic approach to manipulate the Q-factors of the resonances in a Huygens' metasurface over a wide range, and theoretically, in the extreme case, can even approach infinity, i.e.  Huygens' BICs (Figure~\ref{fig:concept}b). Unlike conventional quasi-BICs that are manifested as sharp amplitude modulation in the spectrum, the quasi-BICs operated under the Huygens' condition are manifested as highly dispersive transmission phase modulation over a $2\pi$ range.  This novel phenomenon of Huygens' quasi-BICs (Figure~\ref{fig:concept}b) occurs when two quasi-BICs with matched Q-factors but opposite mode parities are spectrally overlapped. We dub this type of Huygens' metasurfaces \emph{Extreme Huygens' Metasurfaces}. There is no obvious way to realize this novel feature from previous studies of high-Q metasurfaces, since nearly all studied platforms have low-symmetry units and rely on the interaction of so-called bright-modes and dark-modes~\cite{miroshnichenko2010fano,luk2010fano,yang2014all,wu2014spectrally,campione2016broken}. This behavior inevitably introduces an additional resonant background in the spectrum, which is not desirable for Huygens' metasurfaces that require a ``background-free" transmission spectrum. The approach introduced in this letter do not rely on the nearfield coupling of bright modes and dark modes, and it allows a direct tuning of the ``darkness" of the quasi-BICs while maintaining a clean transmission background, offering an improved pathway towards high-Q metasurfaces with high transmission.  

\section{Results and discussion}
\begin{figure*}[t!]
\includegraphics[width=2.05\columnwidth]{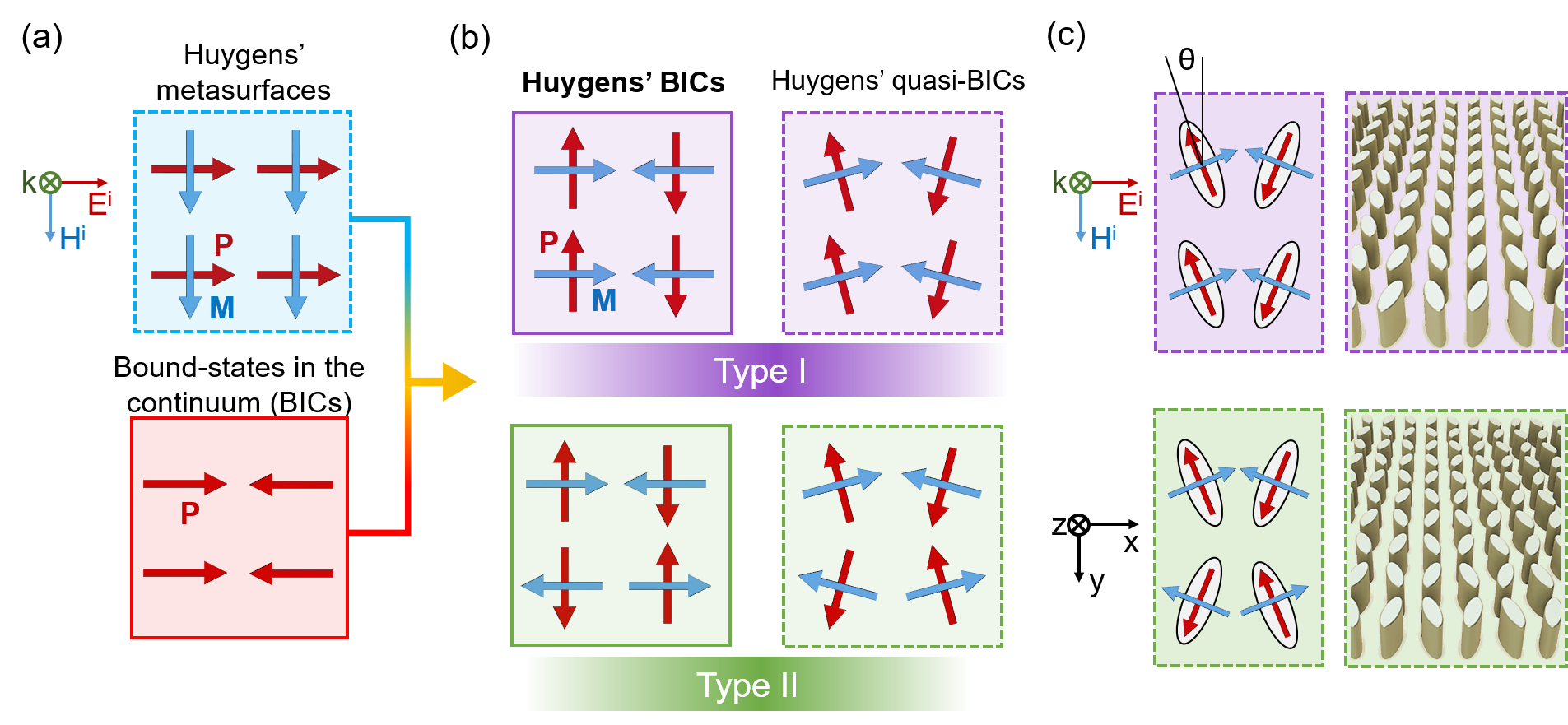}
\caption{ Schematics of the concepts of different states. (a) Conventional Huygens' metasurfaces and optical bound-states in the continuum (BICs). $P$ and $M$ denote the electric and magnetic polarizations, respectively. (b) The concept of Huygens' BICs. Type I and type II correspond to Huygens' BICs of different in-plane symmetry. By introducing structural perturbation, the BICs become quasi-BICs that couple to the scattering channels. (c) Proposed implementation based on zig-zag arrays of anisotropic meta-atoms. The solid (dashed) frame indicates that the state is a bound-state (leaky-state).  \label{fig:concept}}
\end{figure*}

We propose to realize this novel effect using an all-dielectric metasurface composed of zig-zag arrays of anisotropic meta-atoms (see Figure~\ref{fig:concept}c), where the quasi-BICs are the collective resonances built on the Mie-like multipolar modes of each individual meta-atom. This zig-zag design was first introduced for realizing optomechanically-induced chirality~\cite{liu2016polarization}, and has been recently adapted for surface-enhanced mid-infrared molecular fingerprint detection~\cite{tittl2018imaging}. By fine-tuning the geometries including the orientation angle $\theta$, this zig-zag design allows us to precisely control the overlapping of the incident field and the collective modes, without introducing additional resonant background in the transmission spectrum. 

For simplicity, we assume that the zig-zag metasurface is periodic on the $x-y$ plane and positioned in a homogeneous background (Figure~\ref{fig:concept}c). We focus on the optical response when the field is incident and scattered in the normal direction $z$. The forward and backward scattered fields $\mathbf{E}^{\rm f}$ and $\mathbf{E}^{\rm b}$ are contributed by two collective resonances, with effective electric and magnetic polarizations $\mathbf {P}(\omega)$ and $\mathbf {M}(\omega)$, respectively:
\begin{eqnarray}\label{eq:scatfx}
\mathbf{E}^{\rm f}&=&\mathbf{E}^{\rm i}+\frac{{\rm i}\omega\eta }{2}( \mathbf {P}+\frac{1}{c}\mathbf {M}),\\
\mathbf{E}^{\rm b}&=&\frac{{\rm i}\omega\eta }{2}( \mathbf {P}-\frac{1}{c}\mathbf {M}).\label{eq:scatbx}
\end{eqnarray}
Here $\eta=\sqrt{\mu/\epsilon}$ is the wave impedance of the background; $\mathbf {P}= \Sigma_{n=1}^N\mathbf {P}_n$ and $\mathbf {M}=\Sigma_{n=1}^N\mathbf {M}_n$ are the total polarizations within each unit cell; N is the number of meta-atoms in each unit-cell and $n$ denotes the contribution from the $n_{\rm th}$ meta-atom. Typically, the electric polarization $\mathbf {P}$ is contributed by modes with even polarity of electric fields along $z$ direction, while the magnetic polarization $\mathbf {M}$ is contributed by modes with odd polarity along the $z$ direction. For the elliptical-shaped meta-atoms, the dominant electric (magnetic) polarization is along the long (short) elliptical axis of the meta-atoms, and thus $|P_{n,x}/P_{n,y}|\approx\tan\theta$, $|M_{n,x}/M_{n,y}|\approx\cot\theta$. 

The mode profile (polarization current) of the meta-atom can be expanded as a series of electric and magnetic multipoles~\cite{evlyukhin2013multipole}. By comparing the forward and backward scattered fields from an array of multipoles (see Supporting Information) and Eqs.~(\ref{eq:scatfx}) and (\ref{eq:scatbx}), we can connect the effective electric and magnetic polarizations with the multipoles as:
\begin{eqnarray}\label{eq:multipole_P}
\mathbf{P}&=&\frac{1}{A} (\mathbf {p}+\frac{ik^2}{2\omega}\mathbf{z}\times\hat{M}\mathbf{z}-\frac{k^2}{6}\hat{O}\mathbf{zz}+...),\\
{\mathbf{M}}&=&\frac{c}{A} (\frac{k}{\omega}\mathbf {z}\times\mathbf {m}+\frac{ik}{6}\hat{Q}\mathbf{z}+...),\label{eq:multipole_M}
\end{eqnarray}
where $k$ is the wave vector, and $\mathbf{p}$, $\mathbf{m}$, $\hat{Q}$, $\hat{M}$, $\hat{O}$ are the electric dipole, magnetic dipole, electric quadrupole, magnetic quadrupole, and electric octopole moments, respectively.

Without loss of generality, we assume that the incident field is a plane wave with a linear polarization ($E_x^{\rm i}, H_y^{\rm i}$) (see Figure~\ref{fig:concept}c). Due to the mirror symmetry of the zig-zag structure, this incident polarization can only couple to the collective modes that have net polarizations ($P_x, M_y$). Unlike the conventional Huygens' metasurfaces and BICs shown in Figure~\ref{fig:concept}a, the dominant polarizations within the zig-zag metasurface are perpendicular to the incident fields when $\theta$ is small, i.e. $P_{n,y}\gg P_{n,x}$, $M_{n,x}\gg M_{n,y}$, as shown in  Figure~\ref{fig:concept}b. However, since the dominant polarization components of the meta-atoms within each unit-cell have antisymmetric distribution, their net contributions become zero: ${P}_y= \Sigma_{n=1}^N {P}_{n,y}=0$, $ {M}_x=\Sigma_{n=1}^N {M}_{n,x}=0$, and their out-coupling to the plane wave with ($E_y, H_x$) polarization is forbidden. As such, the zig-zag metasurface provides a novel platform to convert the incident energy to collective modes whose radiative loss is largely suppressed due to the symmetry of the mode profile. In the limit of $\theta\rightarrow 0$, the collective modes becomes BICs that are symmetry protected~\cite{hsu2016bound}.

To shed light on how the Q-factors of the quasi-BICs can be controlled by the geometries, we introduce an effective impedance model to describe the interaction between the collective modes and the scattering channel \cite{liu2012optical,liu2014spontaneous,liu2016polarization}:
\begin{eqnarray}\label{eq:imp}
\left[\begin{array}{cc}
Z_{\mathrm{E}} & 0\\
0 & Z_{\mathrm{M}}
\end{array}\right]\left[\begin{array}{c}
I_{{\rm E}}\\
I_{{\rm M}}
\end{array}\right]=\left[\begin{array}{c}
V_{{\rm E}}\\
V_{{\rm M}}
\end{array}\right], 
\end{eqnarray}
where $Z_{\mathrm {E (M)}}$ is the effective impedance of the collective even (odd) mode with an effective electric (magnetic) polarization $P_x (M_y)$ in each unit cell. For brevity, we will dub these two types of quasi-BICs as electric quasi-BIC (E-QBIC) and magnetic quasi-BIC (M-QBIC) below. $I_{\mathrm {E (M)}}$ is the current amplitude of the mode. The interaction between the modes and the incident field is described by the effective electromotive force, defined by the overlapping of the mode and the incident field: $V_{\mathrm{E(M)}}=\int\mathbf{j}_{\mathrm{E(M)}}\cdot\mathbf{E}^{\rm i}\mathrm{d}^3\mathbf{r}$, where $\mathbf{j}_{\mathrm{E(M)}}(\mathbf{r})$ is the normalized electric current distribution of the excited mode. 

 \begin{figure*}[t!]
\includegraphics[width=2.05\columnwidth]{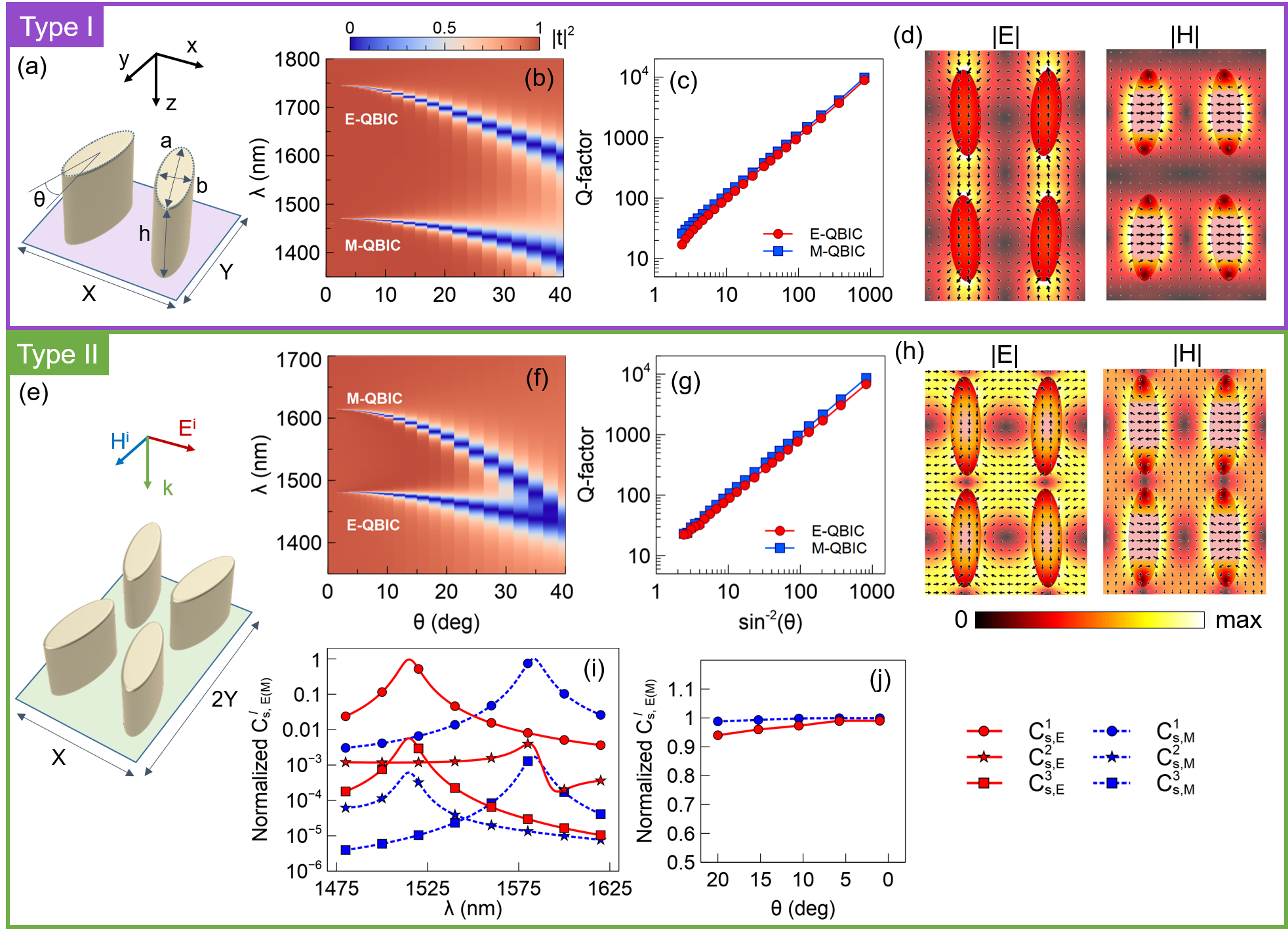}
\caption{ Proposed implementation of electric and magnetic quasi-BICs based on zig-zag arrays of silicon elliptical-cylinders embedded in PMMA. (a) to (d) are the unit-cell structure, simulated transmission spectra, Q-factors and field distributions for type I quasi-BICs, the corresponding results for type II are given from (e) to (h). For the results shown from (b) to (d), $X=900, Y=800, a=550, b=170$, the thickness of meta-atoms is $h=400$; for the results from (f) to (h), $X=900, Y=600, a=530, b=170, h=540$ (units in nm). The E-field plots in (d) and (h) show the time-averaged electric field amplitude and the instantaneous vector E-field of the electric quasi-BICs, at the cross-section plane of $z=h/2$, while the magnetic field plots correspond to fields at the magnetic quasi-BICs. In both plots, $\theta=2^{\circ}$. 
(i) Multipole analysis of the polarization current of a single meta-atom in the array. The scattering cross-sections have been normalized to the peak of the total scattering cross-section. The structural parameters of the metasurface chosen are $\theta=15^{\circ}, X = 900; Y = 650; a = 530; b = 170; h = 540$ (units in nm). (j) The contribution of the two dominant dipole moments at the resonant frequencies under different angles of $\theta$. \label{fig:numeric} }
\end{figure*}

As a representative case, we assume that the dominant contribution of the electric (magnetic) polarization comes from the electric (magnetic) dipole moment of the mode, i.e. the first terms of Eqs.~(\ref{eq:multipole_P}) and (\ref{eq:multipole_M}). Then we have
\begin{eqnarray}\label{eq:V}
[V_{\mathrm{E}},V_{\mathrm{M}}]^{T}=[E_{x}^{{\rm i}}u_{{\rm E}}\sin\theta,ikE_{x}^{{\rm i}}u_{{\rm M}}\sin\theta]^{T}. 
\end{eqnarray}
$u_\mathrm{E(M)}$ is the electric (magnetic) dipole moment of the meta-atom calculated based on the normalized current distribution $\mathbf{j}_{\mathrm{E(M)}}(\mathbf{r})$. Substituting Eq.~(\ref{eq:V}) into Eq.~(\ref{eq:imp}), we can solve the mode amplitudes $I_\mathrm{E(M)}$, and the net polarizations within each unit cell can be expressed explicitly as
\begin{eqnarray}\label{eq:Px}
P_{x}&=&\Sigma_{n=1}^N P_{n,x}=\frac{iNE_{x}^{{\rm i}}u_{E}^{2}}{\omega AZ_{\rm E}} \sin^{2}\theta,\\
M_{y}&=&\Sigma_{n=1}^N M_{n,y}=\frac{iNkE_{x}^{\rm i}u_{M}^{2}}{AZ_{\rm M}}\sin^{2}\theta.\label{eq:My}
\end{eqnarray}
Here, we have used the relations: $P_x=p_x/A=\Sigma_{n=1}^NiI_{\mathrm{E},n}u_{\mathrm E}\sin{\theta}/(\omega A)$, and $M_y=m_y/A=\Sigma_{n=1}^N I_{\mathrm{M},n}u_{\mathrm M}\sin{\theta}/A$, with $p_x$, $m_y$ and $A$ being the net electric dipole, magnetic dipole and the area of each unit cell. Substitute Eqs.~(\ref{eq:Px}) and (\ref{eq:My}) into Eqs.~(\ref{eq:scatfx}) and (\ref{eq:scatbx}), we have the expression of forward and backward scattered fields $E_x^{\rm f}$ and $E_x^{\rm b}$, and the transmission and reflection coefficients are given by:
\begin{eqnarray}\label{eq:t}
t&=&\frac{E_{x}^{\rm f}}{E_{x}^{\rm i}}=1-\frac{N\eta \sin^{2}\theta}{2A}\left(\frac{u_{\rm E}^{2}}{Z_{\rm E}}+\frac{k^{2}u_{\rm M}^{2}}{Z_{\rm M}}\right),\\
r&=&\frac{E_{x}^{\rm b}}{E_{x}^{\rm i}}=-\frac{N\eta \sin^{2}\theta}{2A}\left(\frac{u_{\rm E}^{2}}{Z_{\rm E}}-\frac{k^{2}u_{\rm M}^{2}}{Z_{\rm M}}\right).\label{eq:r}
\end{eqnarray}

Around the resonant frequency, the effective impedance can be expressed as: $Z_{\rm E(M)}=-i\omega L_{\rm E(M)}-1/(i\omega C_{\rm E(M)})+R_{\rm E(M)}$. For systems without material loss, $R_{\rm E(M)}$ represents the effective resistance contributed by the radiative loss, and the total energy should be conserved: $|t|^2+|r|^2=1$. Substituting Eqs.~(\ref{eq:t}) and (\ref{eq:r}) into this requirement leads to the explicit expression for the effective resistance:
\begin{eqnarray}\label{eq:R_E}
R_{\rm E}&=&\frac{N\eta u_{\rm E}^{2}}{2A}\sin^{2}\theta,\\
R_{\rm M}&=&\frac{N\eta k^{2}u_{\rm M}^{2}}{2A}\sin^{2}\theta.\label{eq:R_M}
\end{eqnarray}

The changes of the effective inductances and capacitances are negligible when $\theta$ is small, this indicates that the Q-factors of the collective electric and magnetic modes:
\begin{eqnarray}\label{eq:Q_E}
Q_{\rm E}&=&\sqrt{L_{\rm E}/C_{\rm E}}/R_{\rm E}\propto\sin^{-2}\theta,\\
Q_{\rm M}&=&\sqrt{L_{\rm M}/C_{\rm M}}/R_{\rm M}\propto\sin^{-2}\theta,\label{eq:Q_M}
\end{eqnarray}
both of which grow asymptotically as $\theta$ approaches zero and become infinity when $\theta=0$. This is a direct manifestation of BICs. 

We emphasize that while the analysis above are based on dipole moments, the model can be easily extended by taking into account the higher order multipoles in Eqs.~(\ref{eq:multipole_P}) and (\ref{eq:multipole_M}). It can be proved that the introduction of higher order multipoles does not change the overall physics, and the scaling rule of the Q-factors revealed by Eqs.~(\ref{eq:Q_E}) and (\ref{eq:Q_M}) remains the same, as long as the change of the mode-profile of the meta-atom can be considered as negligible when $\theta$ varies. For more details on the extended model, see Supporting Information.


We first perform a numerical study of a zig-zag metasurface based on hydrogenated amorphous silicon (a-Si:H) meta-atoms embedded in a homogeneous background. To mimic the experimental condition, we employ the experimentally measured permittivity of a-Si:H ($n\approx3.52$ at 1500 nm), and choose Poly(methyl methacrylate)(PMMA) as the background material (refractive index n=1.48). We utilize full-wave simulation (CST Microwave Studio) to design the metasurface such that the electric and magnetic quasi-BICs are located at the near-infrared wavelengths where the material loss of the a-Si:H is negligible (see detailed geometries in Figure~\ref{fig:numeric}). We first design the geometries such that the electric and magnetic quasi-BICs have almost identical Q-factors but remain spectrally separated, so that their individual evolution can be captured clearly. 

We first examine the simplest zig-zag structure shown in Figure~\ref{fig:numeric}), which has mirror symmetry in only one direction. As shown from the twist-angle-dependent transmission spectra in Figure~\ref{fig:numeric}b, the linewidths of both the electric and magnetic quasi-BICs decrease rapidly as $\theta$ approaches zero, while the resonance shifts saturate. Remarkably, their Q-factors, plotted as a function of $\sin^{-2}\theta$ in Figure~\ref{fig:numeric}c, follows an almost-perfect linear relation, confirming the prediction from Eqs.~(\ref{eq:Q_E}) and (\ref{eq:Q_M}). The mode profiles further demonstrate that the two quasi-BICs correspond to collective modes that have antisymmetric distribution of polarization currents, with dominant polarizations perpendicular to the incident one (see Figure~\ref{fig:numeric}d). These two modes can be considered as electromagnetic analogues of the anti-ferromagnetic and anti-ferroelectric states.

The modes in Figure~\ref{fig:numeric}d with an odd parity symmetry along one transverse direction $y$ correspond to the type-I quasi-BICs, as depicted in Figure~\ref{fig:concept}b. This type of quasi-BICs are induced by reducing the mirror symmetry of the structure, which is directly related to various metasurfaces that support ``dark-modes" based on Fano resonances~\cite{Kirill2018Asymmetric}. There are also BICs with odd parity symmetry in  both $x$ and $y$ directions, as depicted by the type-II BICs in Figure~\ref{fig:concept}b. To transform these BICs into observable quasi-BICs, we introduce perturbation to form zig-zag metasurfaces with mirror symmetry in both x and y directions, as shown in Figure~\ref{fig:numeric}e. Although the mode profiles (see Figure~\ref{fig:numeric}h) and the resonant frequencies (see Figure~\ref{fig:numeric}f) are different due to the change of mutual coupling among meta-atoms, their Q-factors still follow the same asymptotic feature described by Eqs.~(\ref{eq:Q_E}) and (\ref{eq:Q_M}), as shown in Figure~\ref{fig:numeric}g.

To show the multipole contribution of the two quasi-BICs studied, we perform a multipole decomposition of the polarization current of a single meta-atom in the array using COMSOL Multiphysics. Following the standard procedure described in \cite{muhlig2011multipole,grahn2012electromagnetic}, we retrieve the coefficients of the vector spherical harmonics $a_E(l,m)$ and $a_M(l,m)$, and calculate the scattering cross-section $C_{s,E(M)}^{l}$ contributed by each order of the electric (magnetic) multipoles (see Figure~\ref{fig:numeric}i as an example). It is clear that the first order moments, i.e. the electric dipole and magnetic dipole are the dominant contribution of the electric and magnetic quasi-BICs studied. Their contribution at the resonant frequencies remains dominant as $\theta$ approaches zero (see Figure~\ref{fig:numeric}j), confirming the assumption of our analytical model. It should be noted that we choose the center of the meta-atom as the origin of the coordinate system for multipole decomposition to minimize unnecessary high order multipoles. 

One interesting question is what will happen if these two quasi-BICs are spectrally overlapped. While the zig-zag array shown in Figure~\ref{fig:numeric}a has a simple unit, the electric polarizations of the meta-atoms along the same column in the y direction oscillates in phase. This bonding effect creates a strong electric near-field enhancement between the tips of the neighboring elliptical meta-atoms. As a result of these ``hot-spots", the electric quasi-BIC becomes very sensitive to the environmental index change and red shifts far-away from the magnetic quasi-BIC due to the presence of PMMA environment (type-I quasi-BICs can be spectrally overlapped and achieve Huygens' quasi-BICs when the environment changes to lower index such as air). To reduce the sensitivity of the electric quasi-BIC and shift it close to the magnetic quasi-BIC, we utilize the zig-zag structure shown in Figure~\ref{fig:numeric}e that supports type-II quasi-BICs. In this design, the electric polarizations in the neighboring meta-atoms along the y direction oscillate out-of-phase (anti-bonding effect), creating ``cold-spots" between the neighboring meta-atoms.

 \begin{figure}[t!]
\includegraphics[width=1.0\columnwidth]{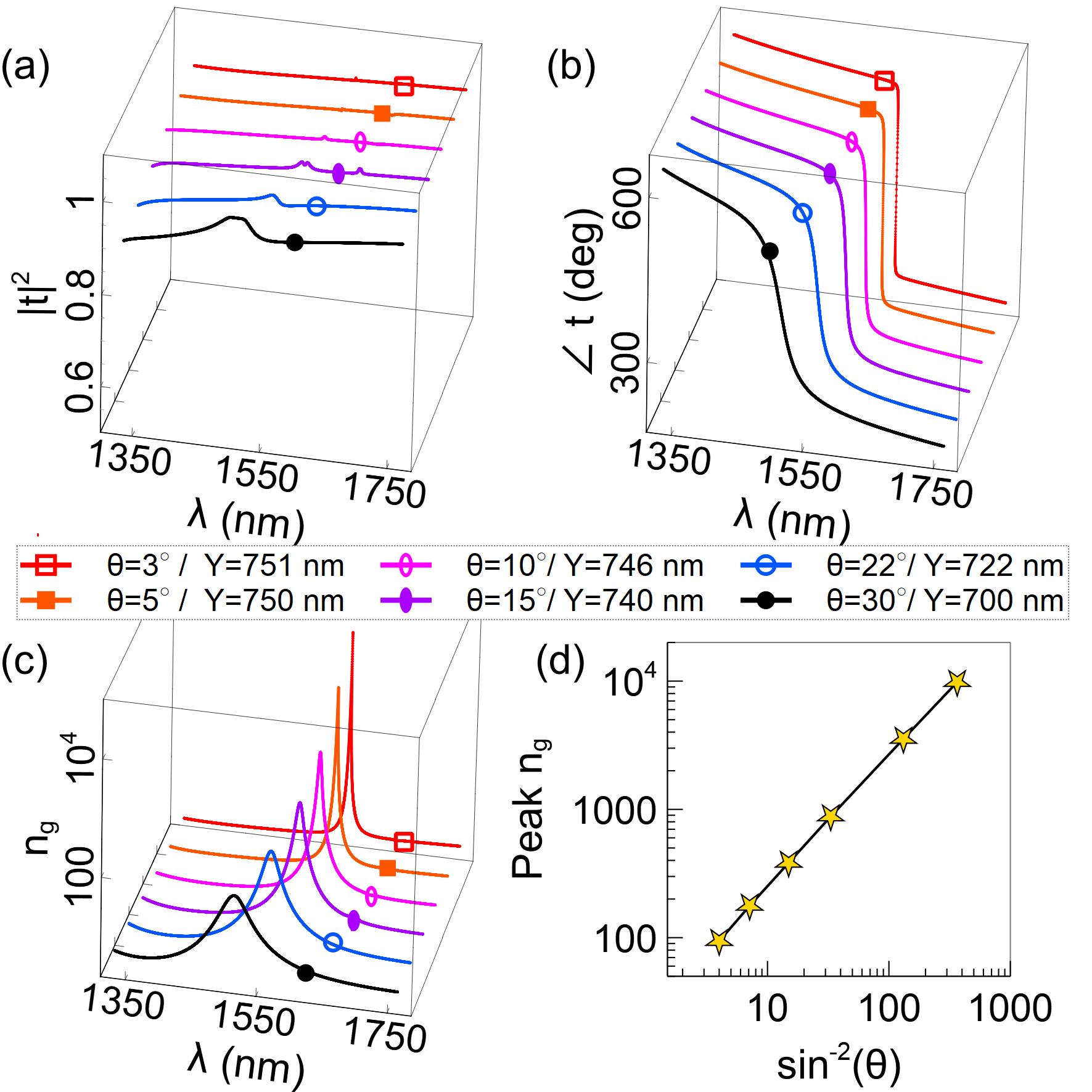}
\caption{ Huygens's metasurfaces with different dispersions. (a) and (b) Simulated transmittance and transmission phase of Huygens' quasi-BICs under different $\theta$ and $Y$. Except for the tuning parameters, all other parameters of the metasurface remain the same as the ones used in Figures~\ref{fig:numeric}e to \ref{fig:numeric}h. (c) The corresponding group index spectra and (d) the peak group index under different $\theta$.   \label{fig:dispersion} }
\end{figure}

By fine-tuning the geometries, the electric quasi-BIC and magnetic quasi-BIC have matching Q-factors and can be spectrally overlapped, and a near unity transmission with a $2\pi$ phase change is realized. In such a case, the transmission can be approximated by the following expression:
\begin{eqnarray}\label{eq:trans}
t\approx 1-\frac{i2\Gamma\omega}{\omega^2-\omega^2_0+i\Gamma\omega}=e^{i\phi(\omega)} 
\end{eqnarray}
$\Gamma=R_{\rm E}/L_{\rm E}=R_{\rm M}/L_{\rm M}$ is the matched linewidth while $\omega_0=\sqrt{1/(L_{\rm E}C_{\rm E})}=\sqrt{1/(L_{\rm M}C_{\rm M}})$ is the matched resonant frequency. 

Although such a Huygens' condition is well-known, the zig-zag metasurface offers an unprecedented capability to control the Q-factors over a wide range while maintaining the Huygens' condition. This is mainly due to the novel property that both $Q_{\rm E}$ and $Q_{\rm M}$ follow the same scaling factor of $\sin^{-2}\theta$ -- once the Q-factors are matched for a particular design, they would remain matching over a wide range of Q as $\theta$ decreases. Figure~\ref{fig:dispersion} (a) and (b) depict the simulated transmittance and phase spectra of a zig-zag Huygens' metasurface with matching electric and magnetic quasi-BICs. By fine-tuning the angle $\theta$ and the lattice constant $Y$ while maintaining other parameters, near full transmission ($>95\%$) with extreme phase dispersion can be realized. 

When Huygens' condition is satisfied, it is physically more meaningful to characterize the dispersion via the effective group index $n_{\rm g}$, which can be evaluated from the group delay $\Delta_{\rm g}=\mathrm{d}\phi(\omega)/\mathrm{d}\omega$:
\begin{eqnarray}\label{eq:groupindex}
n_{\rm g}(\omega)=\frac{c\Delta_{\rm g}}{h}=\frac{c}{h}\frac{2\Gamma(\omega^2+\omega^2_0)}{(\omega^2-\omega^2_0)^2+(\Gamma\omega)^2} 
\end{eqnarray}
where $h$ is the thickness of the metasurface. $n_{\rm g}$ reaches its maximum at $\omega=\omega_0$: 
\begin{eqnarray}\label{eq:maxgroupindex}
n_{\rm g}(\omega_0)=4c/(h\Gamma)=4Qc/(h\omega_0)\propto\sin^{-2}\theta.
\end{eqnarray}
Clearly, the group index follows the same scaling factor as the Q-factor.

As shown in Figure~\ref{fig:dispersion}c, the effective group index can be tuned over several orders of magnitude, and the peak group index follows the same scaling factor as the Q-factor (Figure~\ref{fig:dispersion}d). This novel feature indicates that extreme Huygens' metasurfaces are capable of engineering dispersion over a wide range of bandwidth, which could benefit the development of pulse stretchers and compressors for various pulse-lengths, ranging from femtosecond to nanosecond. Compared to another well-studied mechanism of slow-light based on electromagnetic-induced transparency \cite{yang2014all}, extreme Huygens' metasurfaces avoid the dramatic amplitude change and the superluminal effect, offering a new potential to engineer either a broadband or complicated group-delay spectrum by stacking multilayer of metasurfaces with diverse group-delay responses.   


 \begin{figure*}[t!]
\includegraphics[width=2\columnwidth]{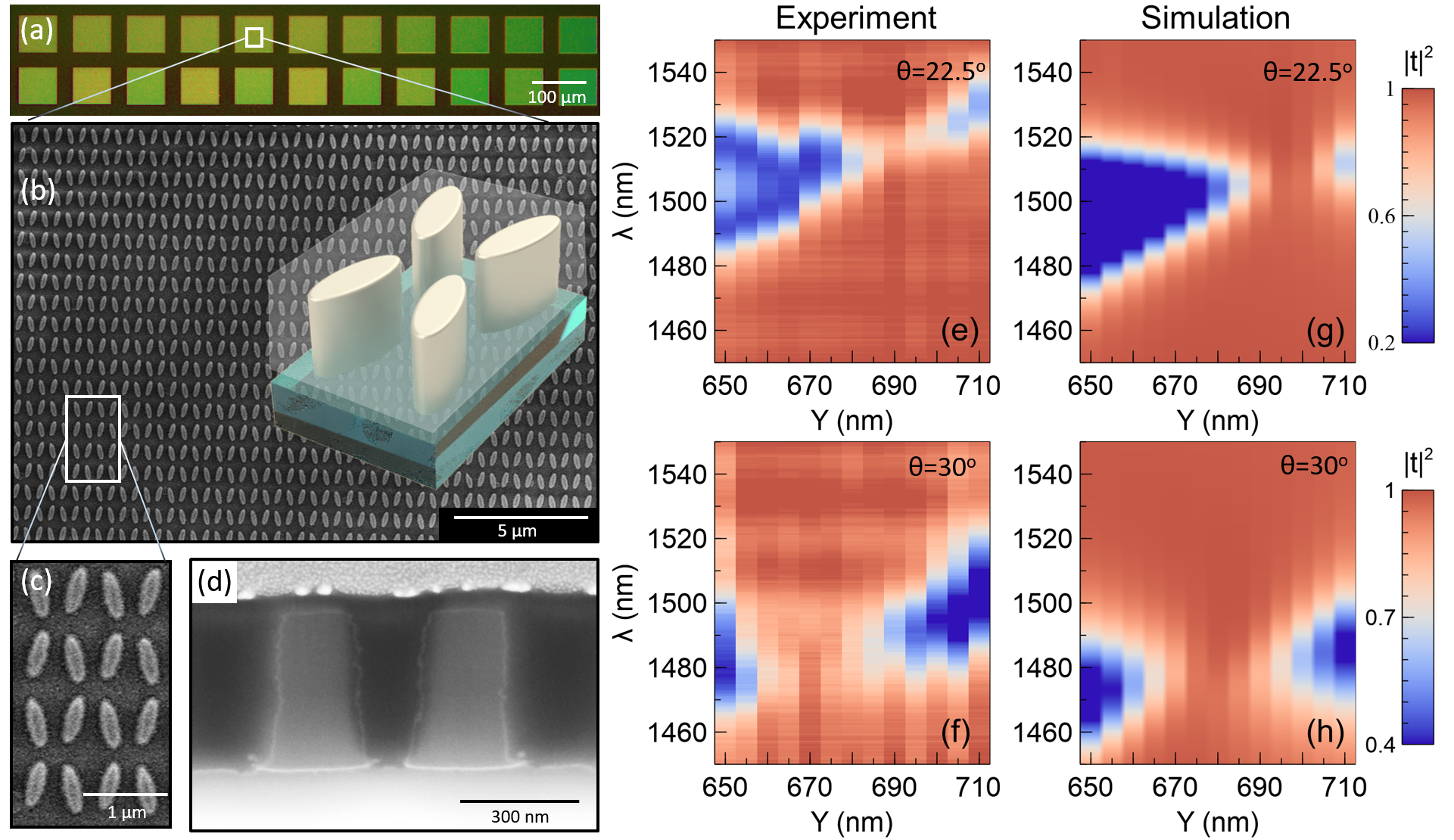}
\caption{Experimental realization of Huygens' quasi-BICs.(a) Optical microscopic images of the fabricated metasurfaces. (b) The SEM images of the metasurfaces. The inset shows the model of the unit-cell covered with an 800-nm PMMA layer. (c) A magnified top view of the  units, and (d) a cross-section of the elliptical cylinders under a 52$^{\circ}$ inclined viewing angle. The measured tapered angle of the elliptical cylinders is around 3.5$^{\circ}$. (e) and (f) The measured transmission spectra and the crossing behavior of the electric and magnetic quasi-BICs under different $Y$ when $\theta=22.5^{\circ}$ and $30^{\circ}$, respectively. (g) and (h) Simulated transmission spectra using post-fabrication measured geometries: $X=900, a=530, b=199, h=520$ (units in nm). \label{fig:exp} }
\end{figure*}

To verify the concept, we fabricated extreme Huygens' metasurfaces working at the near-infrared wavelengths around 1500 nm. The fabrication of the Huygen’s metasurfaces started with cleaning of a slide glass with acetone/IPA/deionized water to promote the adhesion between the substrate and a-Si:H film. Following this, a 520 nm thick a-Si:H was deposited the substrate by plasma-enhanced chemical vapor deposition (Plasmalab 100 from Oxford) using optimized conditions in the previous work~\cite{Gai2014Integrated}. After spin coating of an electron beam resist (ZEP520A from Zeon Chemicals), a thin layer of e-spacer 300Z (Showa Denko) was applied to avoid charging during subsequent electron beam exposure. The metasurface pattern was then formed using an electron beam writer (EBL, Raith150) and subsequent development in ZED-N50. Next, a 50 nm aluminum layer was deposited by e-beam evaporation (Temescal BJD-2000), followed by a lift-off process by soaking the sample in resist remover (ZDMAC from ZEON Co.). The remaining elliptical aluminum pattern array was used as the etch mask to transfer the designed pattern into the a-Si:H film using fluorine-based inductively coupled plasma-reactive ion etching (Oxford Plasmalab System 100). The etching conditions were optimized to obtain a highly anisotropic etching profile for the a-Si:H. Finally, the residual aluminum etch mask was removed by aluminum wet etching solution. 

To minimize the bianisotropy introduced by the glass substrate (n=1.5), a layer of PMMA (n=1.48) of around 800 nm thick was spin-coated to cover the silicon metasurfaces. This thickness was also chosen to minimize the Fabry-Perot reflection around 1500 nm. To show Huygens' metasurfaces with different Q-factors, two designs of samples were fabricated, with $\theta=22.5^{\circ}$ and $ 30^{\circ}$ being the only difference in parameters. For each design, we fabricated a number of arrays (see Figure~\ref{fig:exp}a); each array has a size of around $70{\mu}m\times70{\mu}m$, and the lattice constant $Y$ is fine-tuned with a 5 nm step to overlap the two quasi-BICs.  

 \begin{figure*}[t!]
\includegraphics[width=1.38\columnwidth]{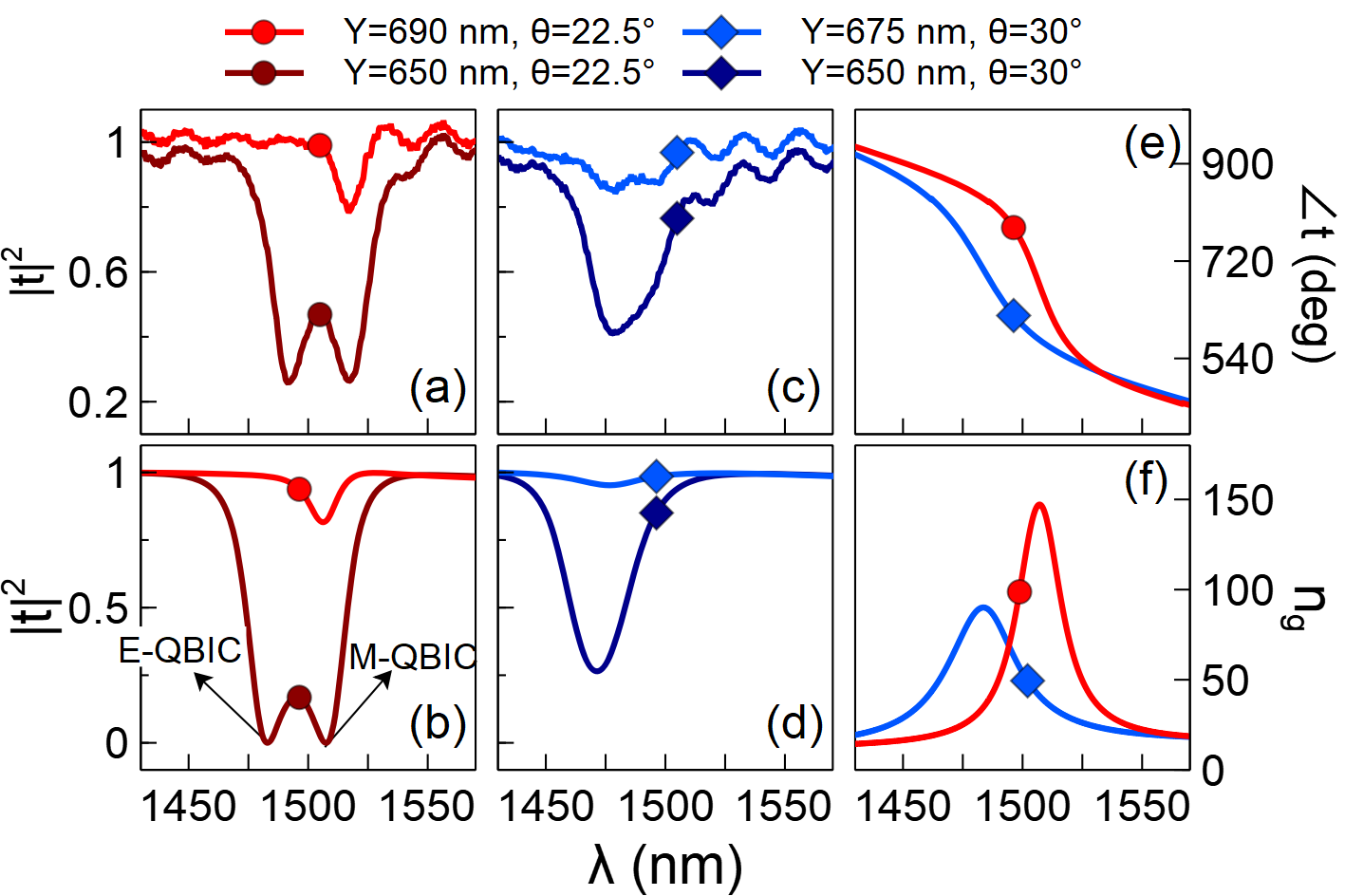}
\caption{A detailed comparison of (a) and (c) the  measured spectra and (b) and (d) the simulated spectra of Huygens' metasurfaces with different designed dispersions. (e) and (f), the corresponding simulated transmission phase and group indexes at the crossing point of electric and magnetic quasi-bound states.  \label{fig:exp2} }
\end{figure*}

The transmission spectra of the samples were measured using a home-built white-light microscopic spectroscopic system. The sample was excited with a linearly-polarized beam with a diverging angle smaller than 2$^{\circ}$, and the signal was collected with a $20\times$ objective. The transmittance spectra were obtained by normalizing the signals of the arrays to that of the unpatterned area. While the Q-factors of the quasi-BICs in the simulation can reach thousands, the maximal Q-factor realized in practice is limited by the imperfection in fabrication such as inhomogeneity, tapering of the cylinders, finite sample size and surface roughness of the meta-atoms. In our experiment, the Q-factor can reach around 300 when the electric and magnetic quasi-BICs are spectrally separated. However, since the electric and magnetic quasi-BICs have different mode profiles, their sensitivity to the imperfection are also different, leading to a small mismatch in the realized Q-factors and additional reflection occurs when the two resonances overlap. The reflection becomes more prominent for high-Q designs due to the increased sensitivity to such imperfection, and therefore the chosen parameters is a trade-off between the Q-factors and the transmission amplitude at the crossing point.    

The measured linewidths of the quasi-BICs are around 20 nm ($Q\approx 75$) and 30 nm ($Q\approx 50$) for the two designs with $\theta=22.5^{\circ}$ and $\theta=30^{\circ}$ (see Figures~\ref{fig:exp}e and \ref{fig:exp}f). As the lattice constant $Y$ increases, the two resonances cross and the resonant transmission increases, which is a direct evidence of the Huygens' effect. Note that since we used a fine tuning step of the parameter $Y$, only designs close to the crossing point are fabricated, and the two resonances are already partially overlapped in the spectra shown in Figure~\ref{fig:exp}f and \ref{fig:exp}h. Using the post-fabrication measured geometries (see captions of Figure~\ref{fig:exp}), we simulate the transmission spectra, and the simulated spectral features agree well with the experimental results except for a small blue-shift of the simulated resonances. 

To have a clearer comparison of the transmission spectra before and at the crossing point of electric and magnetic quasi-BICs, we re-plot two sets of the measured spectra and their corresponding simulated spectra from Figure~\ref{fig:exp}, as shown in Figure~\ref{fig:exp2}a to \ref{fig:exp2}d. At the crossing point, the transmission increases to around 80\%  for design $\theta=22.5^{\circ}$ and 85\% for design $\theta=30^{\circ}$. The simulated transmission phases and the group indexes are given in Figures~\ref{fig:exp2}e and ~\ref{fig:exp2}f, showing that the dispersion of the Huygens' metasurfaces can be fine tuned. The simulated full crossing behavior and the corresponding phase change can be found in Figure S1 of the Supporting Information. From the simulated electromagnetic fields, we confirm that the crossing is indeed from the electric and magnetic quasi-BICs, with the mode profiles similar to the ones shown in Figure~\ref{fig:numeric}d and \ref{fig:numeric}h (see Figure S2 of the Supporting Information for details).

\section{Outlook and Conclusion}

While the design demonstrated in this paper utilized low-order Mie-like resonances, the same effect can also be realized using high-order resonances~\cite{LiuGeneralized}. The elliptical meta-atoms can be substituted with more sophisticated designs, and the perturbation can be introduced in different forms without changing the overall physics~\cite{Kirill2018Asymmetric}. In fact, even the capability of achieving high-Q factors without changing the mode-profiles of the Mie-like resonances is already interesting for many studies such as sensing~\cite{tittl2018imaging} and nonlinear meta-optics~\cite{yang2015nonlinear,lawrence2018nonreciprocal}. The capability of controlling the Q-factors while maintaining Huygens' condition further expand the potential of light-matter interaction at the subwavelength scale. 

For example, it has been shown that the inclusion of highly dispersive materials in an optical cavity can narrow the resonant linewidth dramatically~\cite{soljavcic2005enhancement}, it is therefore particularly interesting to study the behavior when extreme Huygens' metasurfaces interact with conventional cavities based on Fabry-Perot effect. Another potential is for narrow-band perfect absorption with a high-transmission background, which is difficult to realize using conventional approaches in optics. Since our approach allows precise tuning of the radiative decay rates over a dramatic range, it offers a great flexibility to match the nonradiative decay rate ~\cite{asadchy2015broadband,MingDegenerate} even when low-loss materials are employed in the absorber. Apart from the various slow-light related effects discussed above, the current approach can also benefit the development of tunable metadevices with large modulation contrast due to its capability of achieving high-Q resonances~\cite{liu2017ultrathin,komar2018dynamic,howes2018dynamic}. Furthermore, the introduction of spatially varying patterns could lead to meta-devices that are highly spectrally selective, complementing existing metadevices that are relatively broadband and may find applications in optical communication, supercavity lasering~\cite{kodigala2017lasing,rybin2017optical} and compact spectroscopy~\cite{tittl2018imaging}. We further envision that extreme Huygens' metasurfaces could offer an unprecedented platform for the recently-developed time-varying Huygens' metadevices at optical frequencies, with which various novel optical effects unachievable using conventional static platforms can be realized~\cite{liu2018huygens}.

To conclude, we proposed and demonstrated the concept of extreme Huygens' metasurfaces both theoretically and experimentally. We introduced a generic approach to control the quality factors of the resonances in metasurfaces over orders of magnitude while maintaining the parities of their scattered fields. This approach bridges the seemly-unrelated concepts of Huygens' metasurfaces and optical bound-states in the continuum, pointing a practical route to control the dispersion of Huygens' metasurfaces and to realize high-efficiency nano-photonic devices that are highly dispersive and spectrally selective.

\section{Supporting Information} 
Supporting Information Available: Theoretical model with multipole contributions, Simulated spectra and field distributions based on post-fabrication measured geometries.

\section{Acknowledgements}
This work was supported by the Australian Research Council. The authors declare no competing financial interests. The authors thank Prof. Yuri Kivshar for the valuable insight and suggestion. M. Liu owns debt to Prof. Dragomir Neshev and Dr. Andrei Komar for their generous support on optical measurement, and to Dr. Quanlong Yang and Dr. Lei Xu for their  help on COMSOL simulation. Also thanks to Kirill Koshelev, Dr. Sergey Kruk, Dr. David Powell and A/Prof. Ilya Shadrivov for the useful discussions and comments. Thanks to Dr. Li Li from the Australian National Fabrication Facility (ANFF) for the help on SEM and FIB characterizations.

\section{Author contribution}
M. Liu conceived the idea, performed the theoretical and numerical studies as well as the optical measurement; D. Choi fabricated and characterized the morphology of the samples.

\bibliographystyle{articletitle=true}
\bibliography{BIC_HM}

\providecommand{\latin}[1]{#1}
\makeatletter
\providecommand{\doi}
  {\begingroup\let\do\@makeother\dospecials
  \catcode`\{=1 \catcode`\}=2\doi@aux}
\providecommand{\doi@aux}[1]{\endgroup\texttt{#1}}
\makeatother
\providecommand*\mcitethebibliography{\thebibliography}
\csname @ifundefined\endcsname{endmcitethebibliography}
  {\let\endmcitethebibliography\endthebibliography}{}
\begin{mcitethebibliography}{60}
\providecommand*\natexlab[1]{#1}
\providecommand*\mciteSetBstSublistMode[1]{}
\providecommand*\mciteSetBstMaxWidthForm[2]{}
\providecommand*\mciteBstWouldAddEndPuncttrue
  {\def\EndOfBibitem{\unskip.}}
\providecommand*\mciteBstWouldAddEndPunctfalse
  {\let\EndOfBibitem\relax}
\providecommand*\mciteSetBstMidEndSepPunct[3]{}
\providecommand*\mciteSetBstSublistLabelBeginEnd[3]{}
\providecommand*\EndOfBibitem{}
\mciteSetBstSublistMode{f}
\mciteSetBstMaxWidthForm{subitem}{(\alph{mcitesubitemcount})}
\mciteSetBstSublistLabelBeginEnd
  {\mcitemaxwidthsubitemform\space}
  {\relax}
  {\relax}

\bibitem[Pfeiffer and Grbic(2013)Pfeiffer, and Grbic]{pfeiffer2013metamaterial}
Pfeiffer,~C.; Grbic,~A. \emph{Phys. Rev. Lett.} \textbf{2013}, \emph{110},
  197401\relax
\mciteBstWouldAddEndPuncttrue
\mciteSetBstMidEndSepPunct{\mcitedefaultmidpunct}
{\mcitedefaultendpunct}{\mcitedefaultseppunct}\relax
\EndOfBibitem
\bibitem[Liu \latin{et~al.}(2010)Liu, Zhang, Wang, and Jin]{liu2010incident}
Liu,~M.; Zhang,~Y.; Wang,~X.; Jin,~C. \emph{Optics Express} \textbf{2010},
  \emph{18}, 11990--12001\relax
\mciteBstWouldAddEndPuncttrue
\mciteSetBstMidEndSepPunct{\mcitedefaultmidpunct}
{\mcitedefaultendpunct}{\mcitedefaultseppunct}\relax
\EndOfBibitem
\bibitem[Monticone \latin{et~al.}(2013)Monticone, Estakhri, and
  Al{\`u}]{monticone2013full}
Monticone,~F.; Estakhri,~N.~M.; Al{\`u},~A. \emph{Phys. Rev. Lett.}
  \textbf{2013}, \emph{110}, 203903\relax
\mciteBstWouldAddEndPuncttrue
\mciteSetBstMidEndSepPunct{\mcitedefaultmidpunct}
{\mcitedefaultendpunct}{\mcitedefaultseppunct}\relax
\EndOfBibitem
\bibitem[Kim \latin{et~al.}(2014)Kim, Wong, and Eleftheriades]{kim2014optical}
Kim,~M.; Wong,~A.~M.; Eleftheriades,~G.~V. \emph{Physical Review X}
  \textbf{2014}, \emph{4}, 041042\relax
\mciteBstWouldAddEndPuncttrue
\mciteSetBstMidEndSepPunct{\mcitedefaultmidpunct}
{\mcitedefaultendpunct}{\mcitedefaultseppunct}\relax
\EndOfBibitem
\bibitem[Pfeiffer \latin{et~al.}(2014)Pfeiffer, Emani, Shaltout, Boltasseva,
  Shalaev, and Grbic]{pfeiffer2014efficient}
Pfeiffer,~C.; Emani,~N.~K.; Shaltout,~A.~M.; Boltasseva,~A.; Shalaev,~V.~M.;
  Grbic,~A. \emph{Nano letters} \textbf{2014}, \emph{14}, 2491--2497\relax
\mciteBstWouldAddEndPuncttrue
\mciteSetBstMidEndSepPunct{\mcitedefaultmidpunct}
{\mcitedefaultendpunct}{\mcitedefaultseppunct}\relax
\EndOfBibitem
\bibitem[Asadchy \latin{et~al.}(2015)Asadchy, Faniayeu, Ra’Di, Khakhomov,
  Semchenko, and Tretyakov]{asadchy2015broadband}
Asadchy,~V.; Faniayeu,~I.; Ra’Di,~Y.; Khakhomov,~S.; Semchenko,~I.;
  Tretyakov,~S. \emph{Physical Review X} \textbf{2015}, \emph{5}, 031005\relax
\mciteBstWouldAddEndPuncttrue
\mciteSetBstMidEndSepPunct{\mcitedefaultmidpunct}
{\mcitedefaultendpunct}{\mcitedefaultseppunct}\relax
\EndOfBibitem
\bibitem[Epstein \latin{et~al.}(2016)Epstein, Wong, and
  Eleftheriades]{epstein2016cavity}
Epstein,~A.; Wong,~J.~P.; Eleftheriades,~G.~V. \emph{Nature Commun.}
  \textbf{2016}, \emph{7}, 10360\relax
\mciteBstWouldAddEndPuncttrue
\mciteSetBstMidEndSepPunct{\mcitedefaultmidpunct}
{\mcitedefaultendpunct}{\mcitedefaultseppunct}\relax
\EndOfBibitem
\bibitem[Epstein and Eleftheriades(2016)Epstein, and
  Eleftheriades]{PhysRevLett.117.256103}
Epstein,~A.; Eleftheriades,~G.~V. \emph{Phys. Rev. Lett.} \textbf{2016},
  \emph{117}, 256103\relax
\mciteBstWouldAddEndPuncttrue
\mciteSetBstMidEndSepPunct{\mcitedefaultmidpunct}
{\mcitedefaultendpunct}{\mcitedefaultseppunct}\relax
\EndOfBibitem
\bibitem[Wong \latin{et~al.}(2016)Wong, Epstein, and
  Eleftheriades]{wong2016reflectionless}
Wong,~J.~P.; Epstein,~A.; Eleftheriades,~G.~V. \emph{IEEE Antennas Wireless
  Propag. Lett.} \textbf{2016}, \emph{15}, 1293--1296\relax
\mciteBstWouldAddEndPuncttrue
\mciteSetBstMidEndSepPunct{\mcitedefaultmidpunct}
{\mcitedefaultendpunct}{\mcitedefaultseppunct}\relax
\EndOfBibitem
\bibitem[Elsakka \latin{et~al.}(2016)Elsakka, Asadchy, Faniayeu, Tcvetkova, and
  Tretyakov]{elsakka2016multifunctional}
Elsakka,~A.~A.; Asadchy,~V.~S.; Faniayeu,~I.~A.; Tcvetkova,~S.~N.;
  Tretyakov,~S.~A. \emph{IEEE Trans. Antennas Propag.} \textbf{2016},
  \emph{64}, 4266--4276\relax
\mciteBstWouldAddEndPuncttrue
\mciteSetBstMidEndSepPunct{\mcitedefaultmidpunct}
{\mcitedefaultendpunct}{\mcitedefaultseppunct}\relax
\EndOfBibitem
\bibitem[Estakhri and Al{\`u}(2016)Estakhri, and Al{\`u}]{estakhri2016wave}
Estakhri,~N.~M.; Al{\`u},~A. \emph{Phys. Rev. X} \textbf{2016}, \emph{6},
  041008\relax
\mciteBstWouldAddEndPuncttrue
\mciteSetBstMidEndSepPunct{\mcitedefaultmidpunct}
{\mcitedefaultendpunct}{\mcitedefaultseppunct}\relax
\EndOfBibitem
\bibitem[Decker \latin{et~al.}(2015)Decker, Staude, Falkner, Dominguez, Neshev,
  Brener, Pertsch, and Kivshar]{decker2015high}
Decker,~M.; Staude,~I.; Falkner,~M.; Dominguez,~J.; Neshev,~D.~N.; Brener,~I.;
  Pertsch,~T.; Kivshar,~Y.~S. \emph{Adv. Opt. Mater.} \textbf{2015}, \emph{3},
  813--820\relax
\mciteBstWouldAddEndPuncttrue
\mciteSetBstMidEndSepPunct{\mcitedefaultmidpunct}
{\mcitedefaultendpunct}{\mcitedefaultseppunct}\relax
\EndOfBibitem
\bibitem[Arbabi \latin{et~al.}(2015)Arbabi, Horie, Bagheri, and
  Faraon]{arbabi2015dielectric}
Arbabi,~A.; Horie,~Y.; Bagheri,~M.; Faraon,~A. \emph{Nature nanotechnology}
  \textbf{2015}, \emph{10}, 937--943\relax
\mciteBstWouldAddEndPuncttrue
\mciteSetBstMidEndSepPunct{\mcitedefaultmidpunct}
{\mcitedefaultendpunct}{\mcitedefaultseppunct}\relax
\EndOfBibitem
\bibitem[Chong \latin{et~al.}(2015)Chong, Staude, James, Dominguez, Liu,
  Campione, Subramania, Luk, Decker, Neshev, Brener, and
  Kivshar]{chong2015polarization}
Chong,~K.~E.; Staude,~I.; James,~A.; Dominguez,~J.; Liu,~S.; Campione,~S.;
  Subramania,~G.~S.; Luk,~T.~S.; Decker,~M.; Neshev,~D.~N.; Brener,~I.;
  Kivshar,~Y.~S. \emph{Nano Lett.} \textbf{2015}, \emph{15}, 5369--5374\relax
\mciteBstWouldAddEndPuncttrue
\mciteSetBstMidEndSepPunct{\mcitedefaultmidpunct}
{\mcitedefaultendpunct}{\mcitedefaultseppunct}\relax
\EndOfBibitem
\bibitem[Kruk \latin{et~al.}(2016)Kruk, Hopkins, Kravchenko, Miroshnichenko,
  Neshev, and Kivshar]{kruk2016invited}
Kruk,~S.; Hopkins,~B.; Kravchenko,~I.~I.; Miroshnichenko,~A.; Neshev,~D.~N.;
  Kivshar,~Y.~S. \emph{APL Photonics} \textbf{2016}, \emph{1}, 030801\relax
\mciteBstWouldAddEndPuncttrue
\mciteSetBstMidEndSepPunct{\mcitedefaultmidpunct}
{\mcitedefaultendpunct}{\mcitedefaultseppunct}\relax
\EndOfBibitem
\bibitem[Wang \latin{et~al.}(2016)Wang, Kruk, Tang, Li, Kravchenko, Neshev, and
  Kivshar]{wang2016grayscale}
Wang,~L.; Kruk,~S.; Tang,~H.; Li,~T.; Kravchenko,~I.; Neshev,~D.~N.;
  Kivshar,~Y.~S. \emph{Optica} \textbf{2016}, \emph{3}, 1504--1505\relax
\mciteBstWouldAddEndPuncttrue
\mciteSetBstMidEndSepPunct{\mcitedefaultmidpunct}
{\mcitedefaultendpunct}{\mcitedefaultseppunct}\relax
\EndOfBibitem
\bibitem[Khorasaninejad \latin{et~al.}(2016)Khorasaninejad, Chen, Devlin, Oh,
  Zhu, and Capasso]{khorasaninejad2016metalenses}
Khorasaninejad,~M.; Chen,~W.~T.; Devlin,~R.~C.; Oh,~J.; Zhu,~A.~Y.; Capasso,~F.
  \emph{Science} \textbf{2016}, \emph{352}, 1190--1194\relax
\mciteBstWouldAddEndPuncttrue
\mciteSetBstMidEndSepPunct{\mcitedefaultmidpunct}
{\mcitedefaultendpunct}{\mcitedefaultseppunct}\relax
\EndOfBibitem
\bibitem[Kruk and Kivshar(2017)Kruk, and Kivshar]{kruk2017functional}
Kruk,~S.; Kivshar,~Y. \emph{ACS Photonics} \textbf{2017}, \emph{4},
  2638--2649\relax
\mciteBstWouldAddEndPuncttrue
\mciteSetBstMidEndSepPunct{\mcitedefaultmidpunct}
{\mcitedefaultendpunct}{\mcitedefaultseppunct}\relax
\EndOfBibitem
\bibitem[Wang \latin{et~al.}(2018)Wang, Wu, Su, Lai, Chen, Kuo, Chen, Chen,
  Huang, Wang, Lin, Kuan, Li, Wang, Zhu, and Tsai]{wang2018broadband}
Wang,~S. \latin{et~al.}  \emph{Nature nanotechnology} \textbf{2018}, \emph{13},
  227\relax
\mciteBstWouldAddEndPuncttrue
\mciteSetBstMidEndSepPunct{\mcitedefaultmidpunct}
{\mcitedefaultendpunct}{\mcitedefaultseppunct}\relax
\EndOfBibitem
\bibitem[Hsu \latin{et~al.}(2016)Hsu, Zhen, Stone, Joannopoulos, and
  Solja{\v{c}}i{\'c}]{hsu2016bound}
Hsu,~C.~W.; Zhen,~B.; Stone,~A.~D.; Joannopoulos,~J.~D.; Solja{\v{c}}i{\'c},~M.
  \emph{Nature Reviews Materials} \textbf{2016}, \emph{1}, 16048\relax
\mciteBstWouldAddEndPuncttrue
\mciteSetBstMidEndSepPunct{\mcitedefaultmidpunct}
{\mcitedefaultendpunct}{\mcitedefaultseppunct}\relax
\EndOfBibitem
\bibitem[von Neumann and Wigner(1929)von Neumann, and Wigner]{von1929On}
von Neumann,~J.; Wigner,~E. \emph{Phys. Z} \textbf{1929}, \emph{30}, 467\relax
\mciteBstWouldAddEndPuncttrue
\mciteSetBstMidEndSepPunct{\mcitedefaultmidpunct}
{\mcitedefaultendpunct}{\mcitedefaultseppunct}\relax
\EndOfBibitem
\bibitem[Marinica \latin{et~al.}(2008)Marinica, Borisov, and
  Shabanov]{marinica2008bound}
Marinica,~D.; Borisov,~A.; Shabanov,~S. \emph{Physical review letters}
  \textbf{2008}, \emph{100}, 183902\relax
\mciteBstWouldAddEndPuncttrue
\mciteSetBstMidEndSepPunct{\mcitedefaultmidpunct}
{\mcitedefaultendpunct}{\mcitedefaultseppunct}\relax
\EndOfBibitem
\bibitem[Plotnik \latin{et~al.}(2011)Plotnik, Peleg, Dreisow, Heinrich, Nolte,
  Szameit, and Segev]{plotnik2011experimental}
Plotnik,~Y.; Peleg,~O.; Dreisow,~F.; Heinrich,~M.; Nolte,~S.; Szameit,~A.;
  Segev,~M. \emph{Physical review letters} \textbf{2011}, \emph{107},
  183901\relax
\mciteBstWouldAddEndPuncttrue
\mciteSetBstMidEndSepPunct{\mcitedefaultmidpunct}
{\mcitedefaultendpunct}{\mcitedefaultseppunct}\relax
\EndOfBibitem
\bibitem[Lee \latin{et~al.}(2012)Lee, Zhen, Chua, Qiu, Joannopoulos,
  Solja{\v{c}}i{\'c}, and Shapira]{lee2012observation}
Lee,~J.; Zhen,~B.; Chua,~S.-L.; Qiu,~W.; Joannopoulos,~J.~D.;
  Solja{\v{c}}i{\'c},~M.; Shapira,~O. \emph{Physical review letters}
  \textbf{2012}, \emph{109}, 067401\relax
\mciteBstWouldAddEndPuncttrue
\mciteSetBstMidEndSepPunct{\mcitedefaultmidpunct}
{\mcitedefaultendpunct}{\mcitedefaultseppunct}\relax
\EndOfBibitem
\bibitem[Molina \latin{et~al.}(2012)Molina, Miroshnichenko, and
  Kivshar]{molina2012surface}
Molina,~M.~I.; Miroshnichenko,~A.~E.; Kivshar,~Y.~S. \emph{Physical review
  letters} \textbf{2012}, \emph{108}, 070401\relax
\mciteBstWouldAddEndPuncttrue
\mciteSetBstMidEndSepPunct{\mcitedefaultmidpunct}
{\mcitedefaultendpunct}{\mcitedefaultseppunct}\relax
\EndOfBibitem
\bibitem[Hsu \latin{et~al.}(2013)Hsu, Zhen, Lee, Chua, Johnson, Joannopoulos,
  and Solja{\v{c}}i{\'c}]{hsu2013observation}
Hsu,~C.~W.; Zhen,~B.; Lee,~J.; Chua,~S.-L.; Johnson,~S.~G.;
  Joannopoulos,~J.~D.; Solja{\v{c}}i{\'c},~M. \emph{Nature} \textbf{2013},
  \emph{499}, 188\relax
\mciteBstWouldAddEndPuncttrue
\mciteSetBstMidEndSepPunct{\mcitedefaultmidpunct}
{\mcitedefaultendpunct}{\mcitedefaultseppunct}\relax
\EndOfBibitem
\bibitem[Weimann \latin{et~al.}(2013)Weimann, Xu, Keil, Miroshnichenko,
  T{\"u}nnermann, Nolte, Sukhorukov, Szameit, and Kivshar]{weimann2013compact}
Weimann,~S.; Xu,~Y.; Keil,~R.; Miroshnichenko,~A.~E.; T{\"u}nnermann,~A.;
  Nolte,~S.; Sukhorukov,~A.~A.; Szameit,~A.; Kivshar,~Y.~S. \emph{Physical
  review letters} \textbf{2013}, \emph{111}, 240403\relax
\mciteBstWouldAddEndPuncttrue
\mciteSetBstMidEndSepPunct{\mcitedefaultmidpunct}
{\mcitedefaultendpunct}{\mcitedefaultseppunct}\relax
\EndOfBibitem
\bibitem[Monticone and Alu(2014)Monticone, and Alu]{monticone2014embedded}
Monticone,~F.; Alu,~A. \emph{Physical Review Letters} \textbf{2014},
  \emph{112}, 213903\relax
\mciteBstWouldAddEndPuncttrue
\mciteSetBstMidEndSepPunct{\mcitedefaultmidpunct}
{\mcitedefaultendpunct}{\mcitedefaultseppunct}\relax
\EndOfBibitem
\bibitem[Gomis-Bresco \latin{et~al.}(2017)Gomis-Bresco, Artigas, and
  Torner]{gomis2017anisotropy}
Gomis-Bresco,~J.; Artigas,~D.; Torner,~L. \emph{Nature Photonics}
  \textbf{2017}, \emph{11}, 232\relax
\mciteBstWouldAddEndPuncttrue
\mciteSetBstMidEndSepPunct{\mcitedefaultmidpunct}
{\mcitedefaultendpunct}{\mcitedefaultseppunct}\relax
\EndOfBibitem
\bibitem[Kodigala \latin{et~al.}(2017)Kodigala, Lepetit, Gu, Bahari, Fainman,
  and Kant{\'e}]{kodigala2017lasing}
Kodigala,~A.; Lepetit,~T.; Gu,~Q.; Bahari,~B.; Fainman,~Y.; Kant{\'e},~B.
  \emph{Nature} \textbf{2017}, \emph{541}, 196\relax
\mciteBstWouldAddEndPuncttrue
\mciteSetBstMidEndSepPunct{\mcitedefaultmidpunct}
{\mcitedefaultendpunct}{\mcitedefaultseppunct}\relax
\EndOfBibitem
\bibitem[Bulgakov and Maksimov(2017)Bulgakov, and Maksimov]{bulgakov2017light}
Bulgakov,~E.~N.; Maksimov,~D.~N. \emph{Optics Express} \textbf{2017},
  \emph{25}, 14134--14147\relax
\mciteBstWouldAddEndPuncttrue
\mciteSetBstMidEndSepPunct{\mcitedefaultmidpunct}
{\mcitedefaultendpunct}{\mcitedefaultseppunct}\relax
\EndOfBibitem
\bibitem[Rybin \latin{et~al.}(2017)Rybin, Koshelev, Sadrieva, Samusev,
  Bogdanov, Limonov, and Kivshar]{rybin2017high}
Rybin,~M.~V.; Koshelev,~K.~L.; Sadrieva,~Z.~F.; Samusev,~K.~B.;
  Bogdanov,~A.~A.; Limonov,~M.~F.; Kivshar,~Y.~S. \emph{Physical review
  letters} \textbf{2017}, \emph{119}, 243901\relax
\mciteBstWouldAddEndPuncttrue
\mciteSetBstMidEndSepPunct{\mcitedefaultmidpunct}
{\mcitedefaultendpunct}{\mcitedefaultseppunct}\relax
\EndOfBibitem
\bibitem[Zhen \latin{et~al.}(2014)Zhen, Hsu, Lu, Stone, and
  Solja{\v{c}}i{\'c}]{zhen2014topological}
Zhen,~B.; Hsu,~C.~W.; Lu,~L.; Stone,~A.~D.; Solja{\v{c}}i{\'c},~M.
  \emph{Physical review letters} \textbf{2014}, \emph{113}, 257401\relax
\mciteBstWouldAddEndPuncttrue
\mciteSetBstMidEndSepPunct{\mcitedefaultmidpunct}
{\mcitedefaultendpunct}{\mcitedefaultseppunct}\relax
\EndOfBibitem
\bibitem[Bulgakov and Maksimov(2017)Bulgakov, and
  Maksimov]{bulgakov2017topological}
Bulgakov,~E.~N.; Maksimov,~D.~N. \emph{Physical review letters} \textbf{2017},
  \emph{118}, 267401\relax
\mciteBstWouldAddEndPuncttrue
\mciteSetBstMidEndSepPunct{\mcitedefaultmidpunct}
{\mcitedefaultendpunct}{\mcitedefaultseppunct}\relax
\EndOfBibitem
\bibitem[Doeleman \latin{et~al.}(2018)Doeleman, Monticone, Hollander, Al{\`u},
  and Koenderink]{doeleman2018experimental}
Doeleman,~H.~M.; Monticone,~F.; Hollander,~W.; Al{\`u},~A.; Koenderink,~A.~F.
  \emph{Nature Photonics} \textbf{2018}, \emph{352}, 1190--1194\relax
\mciteBstWouldAddEndPuncttrue
\mciteSetBstMidEndSepPunct{\mcitedefaultmidpunct}
{\mcitedefaultendpunct}{\mcitedefaultseppunct}\relax
\EndOfBibitem
\bibitem[Koshelev \latin{et~al.}()Koshelev, Lepeshov, Liu, Bogdanov, and
  Kivshar]{Kirill2018Asymmetric}
Koshelev,~K.; Lepeshov,~S.; Liu,~M.; Bogdanov,~A.; Kivshar,~Y. \emph{Phys. Rev.
  Lett.} \emph{121}, 193903\relax
\mciteBstWouldAddEndPuncttrue
\mciteSetBstMidEndSepPunct{\mcitedefaultmidpunct}
{\mcitedefaultendpunct}{\mcitedefaultseppunct}\relax
\EndOfBibitem
\bibitem[Miroshnichenko \latin{et~al.}(2010)Miroshnichenko, Flach, and
  Kivshar]{miroshnichenko2010fano}
Miroshnichenko,~A.~E.; Flach,~S.; Kivshar,~Y.~S. \emph{Reviews of Modern
  Physics} \textbf{2010}, \emph{82}, 2257\relax
\mciteBstWouldAddEndPuncttrue
\mciteSetBstMidEndSepPunct{\mcitedefaultmidpunct}
{\mcitedefaultendpunct}{\mcitedefaultseppunct}\relax
\EndOfBibitem
\bibitem[Luk'yanchuk \latin{et~al.}(2010)Luk'yanchuk, Zheludev, Maier, Halas,
  Nordlander, Giessen, and Chong]{luk2010fano}
Luk'yanchuk,~B.; Zheludev,~N.~I.; Maier,~S.~A.; Halas,~N.~J.; Nordlander,~P.;
  Giessen,~H.; Chong,~C.~T. \emph{Nature materials} \textbf{2010}, \emph{9},
  707\relax
\mciteBstWouldAddEndPuncttrue
\mciteSetBstMidEndSepPunct{\mcitedefaultmidpunct}
{\mcitedefaultendpunct}{\mcitedefaultseppunct}\relax
\EndOfBibitem
\bibitem[Yang \latin{et~al.}(2014)Yang, Kravchenko, Briggs, and
  Valentine]{yang2014all}
Yang,~Y.; Kravchenko,~I.~I.; Briggs,~D.~P.; Valentine,~J. \emph{Nature
  communications} \textbf{2014}, \emph{5}, 5753\relax
\mciteBstWouldAddEndPuncttrue
\mciteSetBstMidEndSepPunct{\mcitedefaultmidpunct}
{\mcitedefaultendpunct}{\mcitedefaultseppunct}\relax
\EndOfBibitem
\bibitem[Wu \latin{et~al.}(2014)Wu, Arju, Kelp, Fan, Dominguez, Gonzales,
  Tutuc, Brener, and Shvets]{wu2014spectrally}
Wu,~C.; Arju,~N.; Kelp,~G.; Fan,~J.~A.; Dominguez,~J.; Gonzales,~E.; Tutuc,~E.;
  Brener,~I.; Shvets,~G. \emph{Nature communications} \textbf{2014}, \emph{5},
  3892\relax
\mciteBstWouldAddEndPuncttrue
\mciteSetBstMidEndSepPunct{\mcitedefaultmidpunct}
{\mcitedefaultendpunct}{\mcitedefaultseppunct}\relax
\EndOfBibitem
\bibitem[Campione \latin{et~al.}(2016)Campione, Liu, Basilio, Warne, Langston,
  Luk, Wendt, Reno, Keeler, Brener, and Sinclair]{campione2016broken}
Campione,~S.; Liu,~S.; Basilio,~L.~I.; Warne,~L.~K.; Langston,~W.~L.;
  Luk,~T.~S.; Wendt,~J.~R.; Reno,~J.~L.; Keeler,~G.~A.; Brener,~I.;
  Sinclair,~M.~B. \emph{ACS Photonics} \textbf{2016}, \emph{3},
  2362--2367\relax
\mciteBstWouldAddEndPuncttrue
\mciteSetBstMidEndSepPunct{\mcitedefaultmidpunct}
{\mcitedefaultendpunct}{\mcitedefaultseppunct}\relax
\EndOfBibitem
\bibitem[Liu \latin{et~al.}(2017)Liu, Powell, Guo, Shadrivov, and
  Kivshar]{liu2016polarization}
Liu,~M.; Powell,~D.~A.; Guo,~R.; Shadrivov,~I.~V.; Kivshar,~Y.~S. \emph{Adv.
  Opt. Mater.} \textbf{2017}, \emph{5}, 1600760\relax
\mciteBstWouldAddEndPuncttrue
\mciteSetBstMidEndSepPunct{\mcitedefaultmidpunct}
{\mcitedefaultendpunct}{\mcitedefaultseppunct}\relax
\EndOfBibitem
\bibitem[Tittl \latin{et~al.}(2018)Tittl, Leitis, Liu, Yesilkoy, Choi, Neshev,
  Kivshar, and Altug]{tittl2018imaging}
Tittl,~A.; Leitis,~A.; Liu,~M.; Yesilkoy,~F.; Choi,~D.-Y.; Neshev,~D.~N.;
  Kivshar,~Y.~S.; Altug,~H. \emph{Science} \textbf{2018}, \emph{360},
  1105--1109\relax
\mciteBstWouldAddEndPuncttrue
\mciteSetBstMidEndSepPunct{\mcitedefaultmidpunct}
{\mcitedefaultendpunct}{\mcitedefaultseppunct}\relax
\EndOfBibitem
\bibitem[Evlyukhin \latin{et~al.}(2013)Evlyukhin, Reinhardt, Evlyukhin, and
  Chichkov]{evlyukhin2013multipole}
Evlyukhin,~A.~B.; Reinhardt,~C.; Evlyukhin,~E.; Chichkov,~B.~N. \emph{JOSA B}
  \textbf{2013}, \emph{30}, 2589--2598\relax
\mciteBstWouldAddEndPuncttrue
\mciteSetBstMidEndSepPunct{\mcitedefaultmidpunct}
{\mcitedefaultendpunct}{\mcitedefaultseppunct}\relax
\EndOfBibitem
\bibitem[Liu \latin{et~al.}(2012)Liu, Powell, Shadrivov, and
  Kivshar]{liu2012optical}
Liu,~M.; Powell,~D.~A.; Shadrivov,~I.~V.; Kivshar,~Y.~S. \emph{Appl. Phys.
  Lett.} \textbf{2012}, \emph{100}, 111114\relax
\mciteBstWouldAddEndPuncttrue
\mciteSetBstMidEndSepPunct{\mcitedefaultmidpunct}
{\mcitedefaultendpunct}{\mcitedefaultseppunct}\relax
\EndOfBibitem
\bibitem[Liu \latin{et~al.}(2014)Liu, Powell, Shadrivov, Lapine, and
  Kivshar]{liu2014spontaneous}
Liu,~M.; Powell,~D.; Shadrivov,~I.; Lapine,~M.; Kivshar,~Y. \emph{Nature
  Commun.} \textbf{2014}, \emph{5}, 4441\relax
\mciteBstWouldAddEndPuncttrue
\mciteSetBstMidEndSepPunct{\mcitedefaultmidpunct}
{\mcitedefaultendpunct}{\mcitedefaultseppunct}\relax
\EndOfBibitem
\bibitem[M{\"u}hlig \latin{et~al.}(2011)M{\"u}hlig, Menzel, Rockstuhl, and
  Lederer]{muhlig2011multipole}
M{\"u}hlig,~S.; Menzel,~C.; Rockstuhl,~C.; Lederer,~F. \emph{Metamaterials}
  \textbf{2011}, \emph{5}, 64--73\relax
\mciteBstWouldAddEndPuncttrue
\mciteSetBstMidEndSepPunct{\mcitedefaultmidpunct}
{\mcitedefaultendpunct}{\mcitedefaultseppunct}\relax
\EndOfBibitem
\bibitem[Grahn \latin{et~al.}(2012)Grahn, Shevchenko, and
  Kaivola]{grahn2012electromagnetic}
Grahn,~P.; Shevchenko,~A.; Kaivola,~M. \emph{New Journal of Physics}
  \textbf{2012}, \emph{14}, 093033\relax
\mciteBstWouldAddEndPuncttrue
\mciteSetBstMidEndSepPunct{\mcitedefaultmidpunct}
{\mcitedefaultendpunct}{\mcitedefaultseppunct}\relax
\EndOfBibitem
\bibitem[Gai \latin{et~al.}(2014)Gai, Choi, and
  Luther-Davies]{Gai2014Integrated}
Gai,~X.; Choi,~D.-Y.; Luther-Davies,~B. \emph{Opt. Express} \textbf{2014},
  \emph{22}, 9948--9958\relax
\mciteBstWouldAddEndPuncttrue
\mciteSetBstMidEndSepPunct{\mcitedefaultmidpunct}
{\mcitedefaultendpunct}{\mcitedefaultseppunct}\relax
\EndOfBibitem
\bibitem[Liu and Kivshar(2018)Liu, and Kivshar]{LiuGeneralized}
Liu,~W.; Kivshar,~Y.~S. \emph{Opt. Express} \textbf{2018}, \emph{26},
  13085--13105\relax
\mciteBstWouldAddEndPuncttrue
\mciteSetBstMidEndSepPunct{\mcitedefaultmidpunct}
{\mcitedefaultendpunct}{\mcitedefaultseppunct}\relax
\EndOfBibitem
\bibitem[Yang \latin{et~al.}(2015)Yang, Wang, Boulesbaa, Kravchenko, Briggs,
  Puretzky, Geohegan, and Valentine]{yang2015nonlinear}
Yang,~Y.; Wang,~W.; Boulesbaa,~A.; Kravchenko,~I.~I.; Briggs,~D.~P.;
  Puretzky,~A.; Geohegan,~D.; Valentine,~J. \emph{Nano letters} \textbf{2015},
  \emph{15}, 7388--7393\relax
\mciteBstWouldAddEndPuncttrue
\mciteSetBstMidEndSepPunct{\mcitedefaultmidpunct}
{\mcitedefaultendpunct}{\mcitedefaultseppunct}\relax
\EndOfBibitem
\bibitem[Lawrence \latin{et~al.}(2018)Lawrence, Barton~III, and
  Dionne]{lawrence2018nonreciprocal}
Lawrence,~M.; Barton~III,~D.~R.; Dionne,~J.~A. \emph{Nano letters}
  \textbf{2018}, \emph{18}, 1104--1109\relax
\mciteBstWouldAddEndPuncttrue
\mciteSetBstMidEndSepPunct{\mcitedefaultmidpunct}
{\mcitedefaultendpunct}{\mcitedefaultseppunct}\relax
\EndOfBibitem
\bibitem[Solja{\v{c}}i{\'c} \latin{et~al.}(2005)Solja{\v{c}}i{\'c}, Lidorikis,
  Hau, and Joannopoulos]{soljavcic2005enhancement}
Solja{\v{c}}i{\'c},~M.; Lidorikis,~E.; Hau,~L.~V.; Joannopoulos,~J.~D.
  \emph{Physical Review E} \textbf{2005}, \emph{71}, 026602\relax
\mciteBstWouldAddEndPuncttrue
\mciteSetBstMidEndSepPunct{\mcitedefaultmidpunct}
{\mcitedefaultendpunct}{\mcitedefaultseppunct}\relax
\EndOfBibitem
\bibitem[Ming \latin{et~al.}(2017)Ming, Liu, Sun, and Padilla]{MingDegenerate}
Ming,~X.; Liu,~X.; Sun,~L.; Padilla,~W.~J. \emph{Opt. Express} \textbf{2017},
  \emph{25}, 24658--24669\relax
\mciteBstWouldAddEndPuncttrue
\mciteSetBstMidEndSepPunct{\mcitedefaultmidpunct}
{\mcitedefaultendpunct}{\mcitedefaultseppunct}\relax
\EndOfBibitem
\bibitem[Liu \latin{et~al.}(2017)Liu, Susli, Silva, Putrino, Kala, Fan, Cole,
  Faraone, Wallace, Padilla, Shadrivov, and Martyniuk]{liu2017ultrathin}
Liu,~M.; Susli,~M.; Silva,~D.; Putrino,~G.; Kala,~H.; Fan,~S.; Cole,~M.;
  Faraone,~L.; Wallace,~V.~P.; Padilla,~W.~J.; Shadrivov,~I.~V.; Martyniuk,~M.
  \emph{Microsystems \& Nanoengineering} \textbf{2017}, \emph{3}, 17033\relax
\mciteBstWouldAddEndPuncttrue
\mciteSetBstMidEndSepPunct{\mcitedefaultmidpunct}
{\mcitedefaultendpunct}{\mcitedefaultseppunct}\relax
\EndOfBibitem
\bibitem[Komar \latin{et~al.}(2018)Komar, Paniagua-Dom{\'\i}nguez,
  Miroshnichenko, Yu, Kivshar, Kuznetsov, and Neshev]{komar2018dynamic}
Komar,~A.; Paniagua-Dom{\'\i}nguez,~R.; Miroshnichenko,~A.; Yu,~Y.~F.;
  Kivshar,~Y.~S.; Kuznetsov,~A.~I.; Neshev,~D. \emph{ACS Photonics}
  \textbf{2018}, \emph{5}, 1742--1748\relax
\mciteBstWouldAddEndPuncttrue
\mciteSetBstMidEndSepPunct{\mcitedefaultmidpunct}
{\mcitedefaultendpunct}{\mcitedefaultseppunct}\relax
\EndOfBibitem
\bibitem[Howes \latin{et~al.}(2018)Howes, Wang, Kravchenko, and
  Valentine]{howes2018dynamic}
Howes,~A.; Wang,~W.; Kravchenko,~I.; Valentine,~J. \emph{Optica} \textbf{2018},
  \emph{5}, 787--792\relax
\mciteBstWouldAddEndPuncttrue
\mciteSetBstMidEndSepPunct{\mcitedefaultmidpunct}
{\mcitedefaultendpunct}{\mcitedefaultseppunct}\relax
\EndOfBibitem
\bibitem[Rybin and Kivshar(2017)Rybin, and Kivshar]{rybin2017optical}
Rybin,~M.; Kivshar,~Y. \emph{Nature} \textbf{2017}, \emph{541}, 164\relax
\mciteBstWouldAddEndPuncttrue
\mciteSetBstMidEndSepPunct{\mcitedefaultmidpunct}
{\mcitedefaultendpunct}{\mcitedefaultseppunct}\relax
\EndOfBibitem
\bibitem[Liu \latin{et~al.}(2018)Liu, Powell, Zarate, and
  Shadrivov]{liu2018huygens}
Liu,~M.; Powell,~D.~A.; Zarate,~Y.; Shadrivov,~I.~V. \emph{Phys. Rev. X}
  \textbf{2018}, \emph{8}, 031077\relax
\mciteBstWouldAddEndPuncttrue
\mciteSetBstMidEndSepPunct{\mcitedefaultmidpunct}
{\mcitedefaultendpunct}{\mcitedefaultseppunct}\relax
\EndOfBibitem
\end{mcitethebibliography}

\section*{}

 \begin{figure*}[t!]
\includegraphics[width=1\columnwidth]{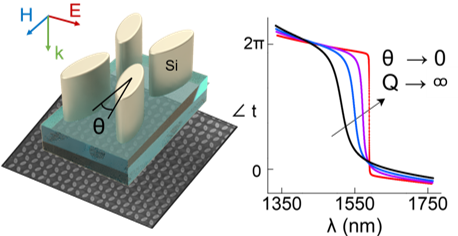}
\end{figure*}

\end{document}